\documentclass[twocolumn,superscriptaddress,preprintnumbers,amsmath,amssymb,floatfix]{revtex4-1}
\pdfoutput=1
\usepackage{graphicx} 
\usepackage{dcolumn} 
\usepackage{bm}
\usepackage{color}
\usepackage{hyperref}
\usepackage{graphicx}
\usepackage{amsmath}
\usepackage{setspace}
\usepackage{appendix}

\begin{document}

\preprint{APS/123-QED}

\title{Extracting neutron skin from elastic proton-nucleus scattering with deep neural network}

\author{G. H. Yang} \thanks{These authors contributed equally to this work.}
\affiliation{School of Physical Science and Technology, Southwest University, Chongqing 400715, China}

\author{Y. Kuang} \thanks{These authors contributed equally to this work.}
\affiliation{School of Physical Science and Technology, Southwest University, Chongqing 400715, China}

\author{Z. X. Yang}\email[zuxing.yang@riken.jp]{}
\affiliation{RIKEN Nishina Center, Wako 351-0198, Japan}
\affiliation{School of Physical Science and  Technology, Southwest University, Chongqing 400715, China}

\author{Z. P. Li}\email[zpliphy@swu.edu.cn]{}
\affiliation{School of Physical Science and Technology, Southwest University, Chongqing 400715, China}

\date{\today}

\begin{abstract}

Based on the relativistic impulse approximation of proton-nucleus elastic scattering theory, the nucleon density distribution and neutron skin thickness of $^{48}$Ca are estimated via the deep learning method. 
The neural-network-generated densities are mainly compressed to be lower inside the nucleus compared with the results from the relativistic PC-PK1 density functional, resulting in a significant improvement on the large-angle scattering observables, both for the differential cross section and analyzing power.
The neutron skin thickness of $^{48}$Ca is captured to be 0.211(11) fm. 
The relatively thicker neutron skin is deemed reasonable from the perspective of density functional analysis. 

\end{abstract}

\maketitle


\section{\label{sec:level1}Introduction}

The equation of state (EOS) plays a crucial role in portraying the underlying basic structure and dynamics of nuclear physics and astrophysics \cite{sym15051123,sym15030683,sym15050994,PhysRevC.102.051303,Shen_2020,HOROWITZ2019167992,epja2020,Thiel_2019}.
Symmetry energy is a key component of the EOS, which quantifies the energy associated with the neutron-proton asymmetry. 
Its density dependence can be gained by the neutron-rich skin of heavy nuclei and the radii of neutron stars \cite{Thiel_2019,PhysRevLett.127.192701}. 
Considering the physical connection between symmetry energy and neutron skin thickness, the measurement of neutron skin becomes crucial in the studies of nuclear many-body dynamics.
To this end, various techniques have been perfected to extract this critical observable. 
Horowitz et al. proposed a direct measurement for neutron skin, where the neutron radius is measured via parity-violating and the proton radius is obtained via elastic electron scattering, respectively \cite{Horowitz_2014}. 
Moreover, coherent pion-photo production \cite{PhysRevLett.112.242502}, measurements of electric dipole polarizabilities \cite{PhysRevC.85.041302,PhysRevLett.107.062502}, etc. are also employed for measuring neutron skins.

Elastic nucleon-nucleus scattering is another important means to study nuclear structure, particularly the medium distributions.
In experimental studies, proton (and ion) elastic scattering is typically performed using inverse kinematics \cite{PhysRevC.92.034608,Krll2016NuclearRI,SAKAGUCHI20171,10.1093/ptep/pty048,PhysRevC.100.054609,Dobrovolsky2019NuclearmatterDI,DOBROVOLSKY2021122154} facilities such as the CSRe storage ring of HIRFL-CSR \cite{XIA200211}, GSI/FAIR \cite{0ab80b2c8f8a4f81970c9c9fbc096c1c}, and the RIBF at RIKEN \cite{10.1063/1.3485213}. 
Exploiting experimental proton elastic scattering data, the nuclear root-mean-square radii and matter density distributions can be extracted by combining theoretical models with approximation methods \cite{PhysRevC.18.2641,PhysRevC.19.1855,PhysRevC.21.1488,PhysRevC.49.2118,PhysRevC.65.044306,PhysRevC.67.054605,KAKI200399,PhysRevC.77.024317,PhysRevC.82.044611,PhysRevC.87.034614,doi:10.1142/S0218301315500159,SAKAGUCHI20171,Zenihiro2018DirectDO,e72caed11f9c4f80bec9086cea8cc196,PhysRevC.107.064310,PhysRevC.108.014614}. 
For example, the matter radii and densities of $^{20, 22}$Ne, $^{24, 26}$Mg at 0.8 GeV \cite{PhysRevC.107.064310}, as well as the $^{78}$Kr at 153 MeV \cite{PhysRevC.108.014614} were obtained most recently based on Glauber multiple-scattering theory and the two-parameter Fermi (2pF) model. 
The neutron skin thickness of $^{48}$Ca and $^{208}$Pb \cite{PhysRevC.65.044306,e72caed11f9c4f80bec9086cea8cc196} were calculated based on Brueckner theory and $g$-matrix folding model. 
The relativistic impulse approximation (RIA), developed by McNeil et al.~\cite{PhysRevLett.50.1439,PhysRevC.27.2123} and Clark et al.~\cite{PhysRevLett.50.1644,PhysRevC.28.1421}, is another practical approach to accurately describe proton elastic scattering \cite{PhysRevLett.50.1439,PhysRevC.27.2123,PhysRevLett.50.1644,PhysRevC.28.1421,PhysRevLett.50.1443,PRCnp1987,PhysRevC.35.1442,PhysRevC.89.014620,PhysRevC.98.014620,PhysRevC.103.064604,sym14030474,PhysRevC.106.034321,etde_20303117,10.1093/ptep/ptx116,PhysRevC.78.014603,PhysRevC.77.014001}, by which the 2pF distribution of $^{9}$C \cite{PhysRevC.87.034614} and the density functional theory derived neutron density distribution of Sn isotopes \cite{PhysRevC.77.024317}, $^{204, 206, 208}$Pb \cite{PhysRevC.82.044611}, and $^{40, 48}$Ca \cite{Zenihiro2018DirectDO} were obtained at energies around 300 MeV. 
Aiming at the above research at 300 MeV, it is essential to modify the relativistic Love-Franey $NN$ interaction to incorporate the effects of the nuclear medium during the RIA-based scattering processes, such as Pauli blocking, multi-step processes, and vacuum polarization effects. 
Furthermore, by systematically studying elastic proton-nucleus scattering, it was found that as the reaction energy transitions from intermediate to high energy, many medium effects, including Pauli blocking, can be neglected, leading to a better-described proton elastic scattering \cite{ky}. 

The studies on the basis of RIA are undoubtedly of great significance, but there are certain non-negligible flaws that persist due to the approximation method in the models, such as 2pF approximation \cite{PhysRevC.87.034614} and the shell contributions with single particle approximation \cite{Zenihiro2018DirectDO}. 
At this juncture, deep learning, as an emerging means, is expected to facilitate overcoming the limitations. 
From the early studies in which machine learning was utilized as a universal approximator \cite{doi:10.1142/S0217979206036053,PhysRevC.53.2358}, 
to Bayesian model averaging (BMA) for more accurate predictions \cite{PhysRevLett.122.062502, PhysRevC.101.044307}, back-propagation neural networks \cite{1986/10/01} have achieved a series of successes in diverse aspects of nuclear complex systems \cite{PhysRevC.101.045204,ATHANASSOPOULOS2004222,GERNOTH19931,Utama_2016,PhysRevC.102.054323,PhysRevC.105.014308,PhysRevC.105.034320,PhysRevLett.124.162502,WANG2022137154,PhysRevC.99.054308,Saxena_2021,PhysRevC.99.064307,PhysRevLett.123.122501,PhysRevC.103.034621}, especially in nuclear mass \cite{PhysRevC.101.045204,ATHANASSOPOULOS2004222,PhysRevLett.122.062502,PhysRevC.101.044307} and nuclear charge radius \cite{Utama_2016,PhysRevC.102.054323,PhysRevC.105.014308,PhysRevC.105.034320}.
On the other hand, generative models are gaining widespread acceptance across diverse fields, with their research focus transitioning from low-dimensional data to large-scale data such as images and texts \cite{bubeck2023sparks}.
Considering the rapid advancements in deep learning, integrating it with RIA theory for calculating the neutron skin thickness would  be a compelling research avenue.

In this study, we will construct a RIA calculation-based generative neural network, referred to as the observable-to-density network (OTDN), to map from observables (differential cross section and analyzing power) to nuclear density distributions.
Building upon this, further integration under Bayesian principles \cite{PhysRevLett.122.062502} will be performed to calculate neutron skin thickness.
The results will be further compared with experiments and other theoretical calculations.

\section{observable-to-density network}\label{sec2}

\begin{figure*}[th]
  \includegraphics[width=17cm]{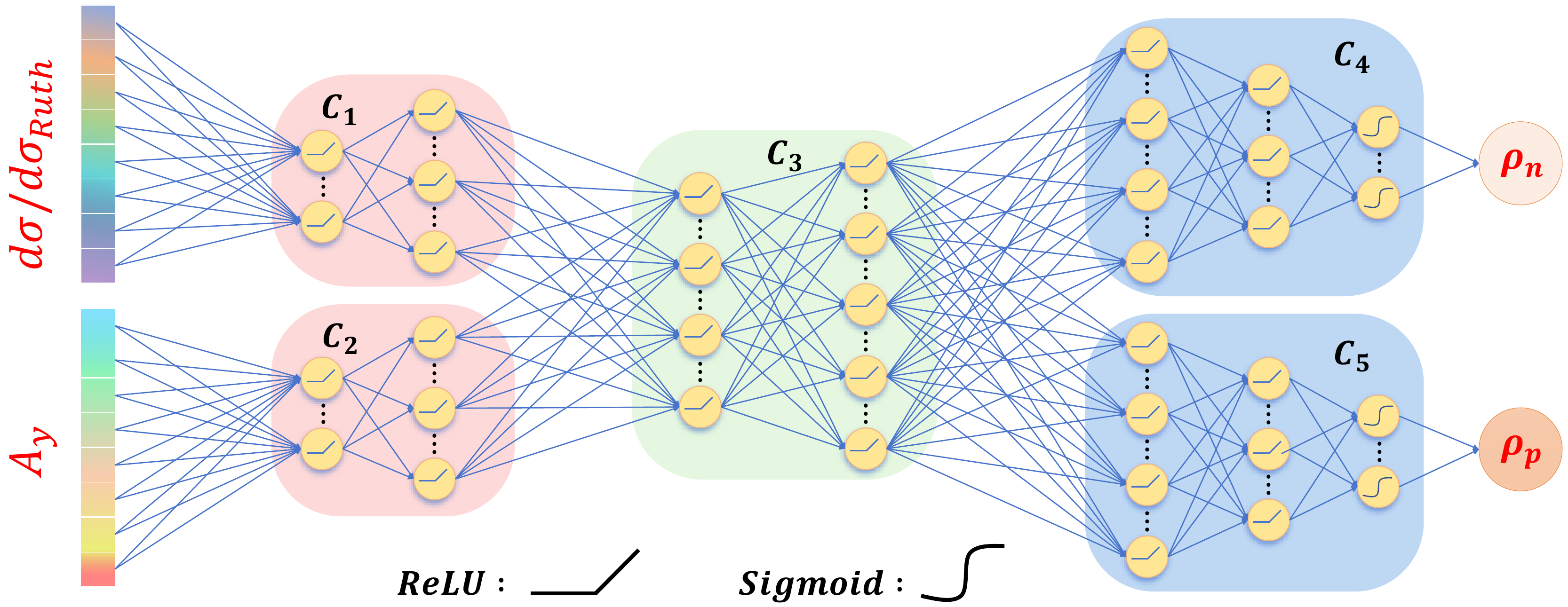}
  \caption{\label{fig:epsart1} The feedforward schematic diagram of observable-to-density network (OTDN).}
\end{figure*}
 
The observables of proton elastic scattering are theoretically obtained by solving the optical potential-based Dirac equation.
The RIA plays a role in obtaining the optical potential by folding densities and $NN~t$-matrices \cite{PhysRevC.35.1442}, written by
\begin{equation}
U_{\mathrm{opt}}(r, E) 
  = \sum_{L}\left[U_{D}^{L}(r, E)+U_{X}^{L}(r, E)\right] \lambda^{L} ,
\end{equation}
with direct term being
\begin{equation}
  U_{D}^{L}(r, E) = -\frac{4 \pi i p}{M} \int \mathrm{d}^{3} r^{\prime} \rho_{L}\left(\mathbf{r}^{\prime}\right) t_{D}^{L}\left(\left|\mathbf{r}^{\prime}-\mathbf{r}\right| ; E\right) ,
\end{equation}
and exchange term being
\begin{align}
    U_{X}^{L}(r, E) =& \notag\\
    - \frac{4 \pi i p}{M} \int \mathrm{d}^{3}& r^{\prime} \rho_{L}\left(\mathbf{r}^{\prime}, \mathbf{r}\right) t_{X}^{L}\left(\left|\mathbf{r}^{\prime}-\mathbf{r}\right| ; E\right) j_{0}\left(p\left|\mathbf{r}^{\prime}-\mathbf{r}\right|\right),
\end{align}
where $\lambda^{L}$ corresponds to the Dirac matrix, $L = S (V)$ as the main focus of optical potential, marks scalar (vector) components of each variable, $j_{0}$ is the spherical Bessel function, and $p$ means the magnitude of the three-momentum of the projectile in the nucleon-nucleus center-of-mass frame. 
Eventually the forward propagation from densities to observables can be expressed as
\begin{equation}
  \{\rho_{L,\tau}, t^L\} \xrightarrow{\text{Folding}} U_{\mathrm{opt}} \xrightarrow{\text{Dirac~equation}} \{ \frac{d\sigma}{d\sigma_\text{Ruth}}, A_{y} \},
\end{equation}
where ${d\sigma}/{d\sigma_\text{Ruth}}$ is the  differential cross sections in the Rutherford ratio (Hereinafter abbreviated as differential cross section) and $A_{y}$ indicates the analyzing power. 

Through the 5-cell OTDN, the inverse mapping of the aforementioned relationship can be obtained, as illustrated in Fig.~\ref{fig:epsart1}.
During the feedforward process, differential cross section $d\sigma /d\sigma_\text{Ruth}$ and analyzing power $A_{y}$ are independently input into two fully connect neural network cells ($C_1$ and $C_2$).
The outputs merge and further non-linear feedback in $C_3$, producing 1024 features.
The massive latent features are fed into $C_4$ and $C_5$, generating the final density distribution.
Each cell consists of several fully connected network layers, and each network layer undergoes non-linear activation. 
Except for the last layer, which uses a Sigmoid ($= 1/(1+e^{-x})$) function for smoothing and constraining the range, all hidden layers are activated through ReLU ($=\max \{0, x\}$).

Aiming to the back-propagation of supervised learning, the loss function based on the normalized flow of Pearson $\chi^2$ divergence (NPD) are employed, written as
\begin{equation}
\text{NPD} = \left\langle \frac{\left[\lambda \rho_{\text {pre }}\left(r\right)-\rho_{\text {tar }}\left(r\right)\right]^{2}}{\lambda \rho_{\text {pre }}\left(r\right)}\right\rangle,
\end{equation}
with
\begin{equation}
    \lambda = \frac{X}{\int_{0}^{\infty} 4 \pi \rho_{\text {pre}}(r) r^{2} d r},
\end{equation}
where $X=Z (N)$ represents the proton (neutron) number.
The factor $\lambda$ normalizes predicted densities, and allow OTDN naturally ignoring overall errors on the order of magnitude and focusing more on the scale change of the shape. 
On the other hand,  the NPD denominator $\lambda \rho_{\text {pre }}\left(r\right)$ weights the error, allowing for a more precise description of the dispersion of surface nucleons.
Thus, the maximum likelihood estimation (MLE) of all network parameters $\textbf{w}$ can be achieved by minimizing the NPD, denoted as
\begin{equation}
\mathbf{w}^{\mathrm{MLE}}=\arg \min _{\mathbf{w}} \text{NPD}(\mathbf{w}).
\end{equation} 

We choose the adaptive moment estimation (Adam) optimizer for the training process. Although Adam automatically adjusts the learning rate, we further attenuate the learning rate to expedite convergence.
Simultaneously, all variables are min-max scaled to enhance training efficiency.
All hyperparameters are listed in supplementary material \cite{SM}.

\section{RESULTS}

Physically, the $NN~t$-matrices utilized in RIA exhibit energy dependence, originating from masses, cutoff parameters, etc. 
Within the 50-200 MeV incident energy range, the parameter sets applied in system calculations are elucidated in Ref.~\cite{ky}, while the parameter sets for 200-500 MeV \cite {MAXWELL1996509} and 500-800 MeV \cite {MAXWELL1998747} incident energy regions are fitted by Maxwell in the late 20th century.
Enhancing the reliability of RIA is a key focus of this work. 
To this end, we opt for a higher energy of 800 MeV, as Pauli blocking and other medium effects have a modest impact on this energy.
Aiming at the energy dependence, the parameters are further optimized via the Levenberg-Marquardt method \cite{Levenberg1944,Marquardt1963}, which are presented in supplementary material \cite{SM}.
Hereinafter, we will denote RIA based on the Maxwell parameter set as ``RIA" and the version of the optimized parameter as ``RIA*". 

To extract detailed information about the target nuclear densities from observables, constructing an appropriate dataset for the target nuclear density is another key focus.
By combining different theoretical calculations to systematically generate a large number of densities so as to obtain additional training samples that cover the experimental observational range comprehensively. 
This approach is technically represented as
\begin{equation}
    \rho_g = \sum_i^D R_i \times \rho_i    
\end{equation}
with the normalized random weight $ \sum_i^D R_i = 1$, where $\rho_i$ represents the elemental densities, $\rho_g$ indicates the generated densities.
The notation $D = 16$ indicates the types of adopted elemental densities, calculated via the relativistic Hartree-Bogoliubov (RHB) with DD-ME2 \cite{PhysRevC.71.024312}, PC-PK1 \cite{PhysRevC.82.054319}, and DD-PC1 \cite{PhysRevC.78.034318} density functionals, the relativistic mean field (RMF) with PK1 density functional \cite{PhysRevC.69.034319}, and the 12 sets of Skyrme parameterizations under the Skyrme-Hartree-Fock (SHF) approach \cite{PhysRevC.33.335}.
In particular, the scalar densities with non-relativistic Skyrme parameterizations are assumed as the linear transformation of vector densities,  
   \begin{equation}\label{1}
    \begin{array}{c}
      \rho_{S}=a_{p{(n)}}\rho_{V},
     \end{array}
   \end{equation}
where $a_{p{(n)}}$ are the linear coefficient of protons (neutrons). 
According to multiple sets of scalar and vector densities, the empirical values $a_{p}=0.961$ and $a_{n}=0.959$ are obtained.
These theories effectively combine models of relativity and non-relativity, covering a sufficiently large range of symmetry energy slopes. 
In this sense, the final densities can reflect a more realistic isospin asymmetry.
For spherical $^{48}\text{Ca}$, only the radial densities are necessary, for which 160 grid points with an interval of 0.1 fm are taken, as nuclei are typically bounded within a radius of 16 fm.
As a consequence, 600 sets of pseudodensities are generated, which have been verified to be adequate for training the network.

\begin{figure}[tb]
  \includegraphics[width=8.5 cm]{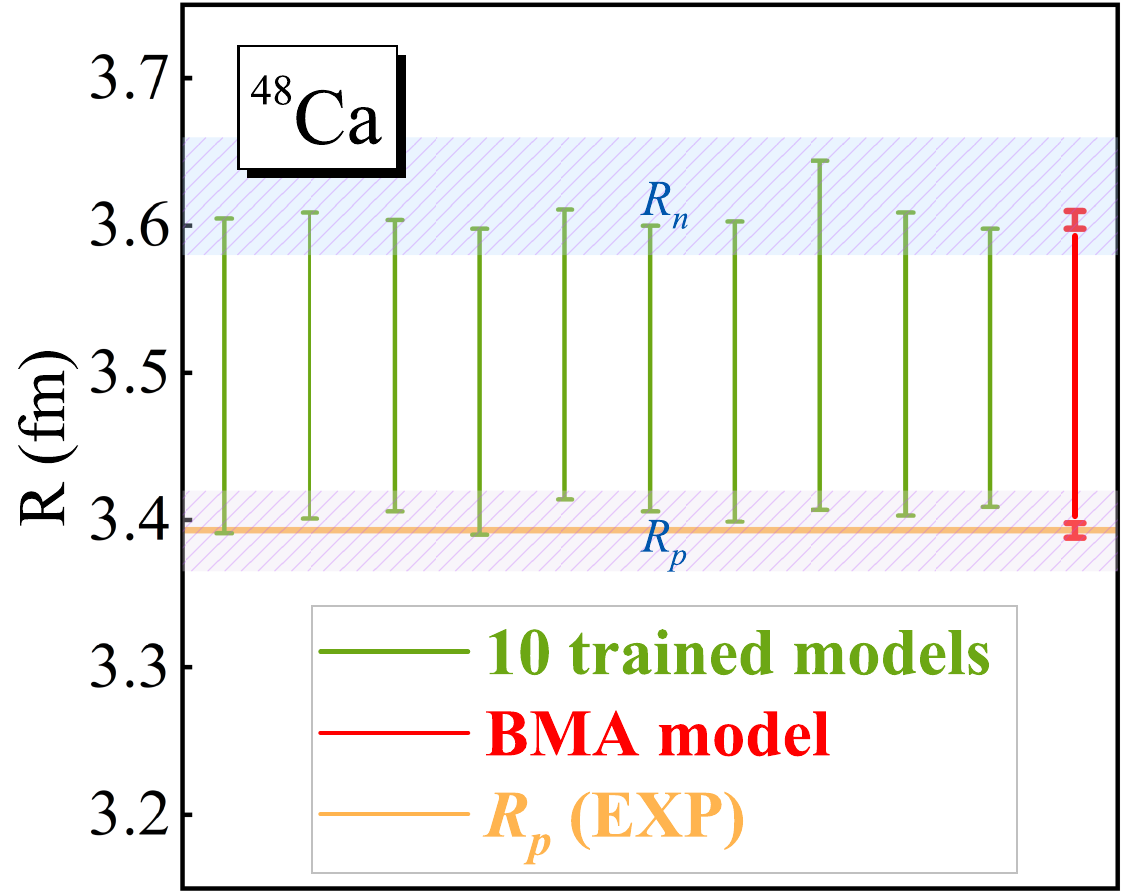}
  \caption{\label{fig:epsart5} The proton radii (falling within the lower shadow) and neutron radii (falling within the upper shadow)  generated by 10 sets of different initializations of OTDN and the posterior results based on bayesian model averaging (BMA).}
\end{figure}

When the experimental observables are input into the trained OTDN, the revised densities are generated.
Nevertheless, during the back-propagation process in search of maximum likelihood, errors caused by stochastic gradient descent need to be further eliminated.
Put differently, the random seed subtly influences the results.
The 10 predictions from different initializations of OTDN are depicted in Fig.~\ref{fig:epsart5}.

Establishing a constraint to limit these predictions is demanding.
Taking the experimentally determined proton radius of $\mathrm{^{48}Ca}$ as prior, we can define weights and obtain posterior predictions of the neutron skin based on the Bayesian principle \cite{PhysRevLett.122.062502}, i.e.,
\begin{align}
w_{k} &:= \mathrm{P}\left(\mathcal{M}_{k} \mid R_{p}^\text{theo} \rightarrow R_{p}^{\exp}\right)  \notag  \\ 
&\propto \mathrm{P}\left(R_{p}^\text{theo} \rightarrow R_{p}^{\exp}  \mid \mathcal{M}_{k}\right) \mathrm{P}\left(\mathcal{M}_{k}\right).
\end{align}
Here, $\mathcal{M}_k$ denotes the trained OTDN, $\mathrm {P(\mathcal{M}_{k})}$ are uniform prior weights, and the arrow represents the proximity between theory and experiment values characterized by square deviation.
The final weights are expressed as 
\begin{equation}
    w_{k}=\frac{\left(R_{p, k}-R_{p}^{\mathrm{Exp}}\right)^{-2}}{\sum_{k=1}^{10}\left(R_{p, k}-R_{p}^{\mathrm{Exp}}\right)^{-2}},
 \end{equation}
where the second power exponent emphasizes the sensitivity to the proton radius.
Phenomenologically, there is a connection between the charge radius $R_{ch}$ measured from electron scattering experiments and the proton radius $R_{p}$,
\begin{equation}
  R_{ch}^{2}=R_{p}^{2}+(0.862 \, \mathrm{fm})^{2}-(0.336 \, \mathrm{fm})^{2} \frac{N}{Z},
\end{equation}
where the second and third terms are determined by the sizes of the proton and neutron, respectively. 
Then, the predicted densities can be expressed as
 \begin{equation}
    \rho_{\tau,\text{final}}=\sum_{k=1}^{10} \rho_{\tau,k} \times w_{k}, 
 \end{equation}
where the $\tau=p,n$ correspond to proton and neutron. 
For the corresponding weighted matter density radii, refer to the red line segment in Fig.~\ref{fig:epsart5}.

 \begin{figure}[tb]
  \includegraphics[width=8.5 cm]{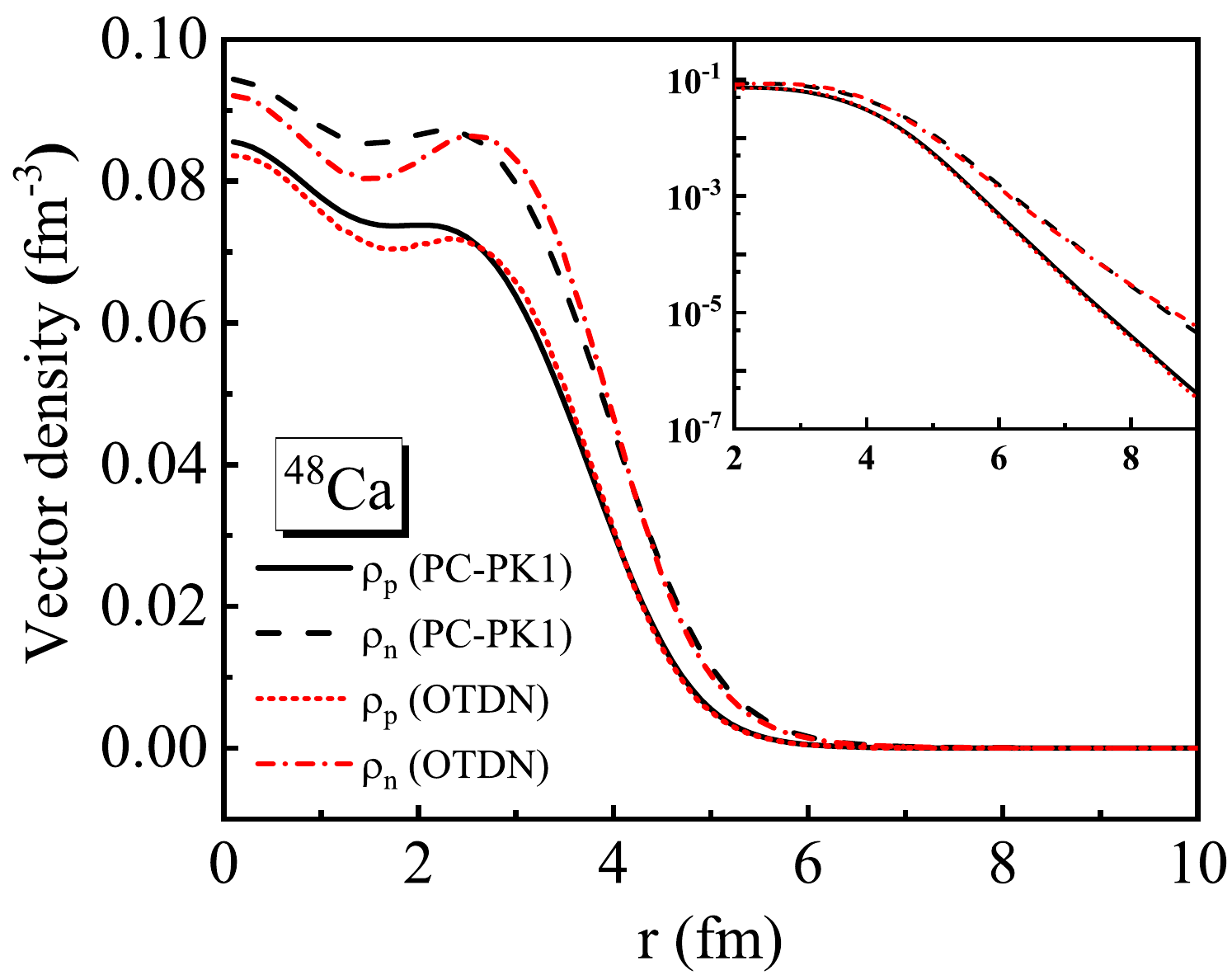}
  \caption{\label{fig:epsart2} Comparison of the PC-PK1 density functional with OTDN for the $\mathrm{^{48}Ca}$ proton and neutron vector densities. 
                The inner panel shows the densities in logarithmic scale.}
\end{figure}

The proton and neutron density distributions of $\mathrm{^{48}Ca}$ extracted by OTDN are shown in Fig.~\ref{fig:epsart2}, which are compared to the PC-PK1 functional calculations.
It can be observed that the OTDN-extracted proton and neutron densities are slightly lower inside the nucleus.
Additionally, according to the inset panel on the logarithm scale, it is clear that the diffuseness at the nuclear surface is consistent.
This implies that the densities near the nuclear surface increases to a certain extent. 
As a consequence, OTDN not only enables the observation of the nuclear surface diffuseness, but it also facilitates the detection of variations in internal density.
Aiming at density functional theory, the new densities leads to variations in various nuclear properties, such as binding energy per nucleon, nuclear shell effects, correlations among nucleons, etc.
These can be calculated through the inverse kohn-sham scheme \cite{inverseks}, warranting further exploration in future endeavors.

\begin{figure}[tb]
  \includegraphics[width=8.5 cm]{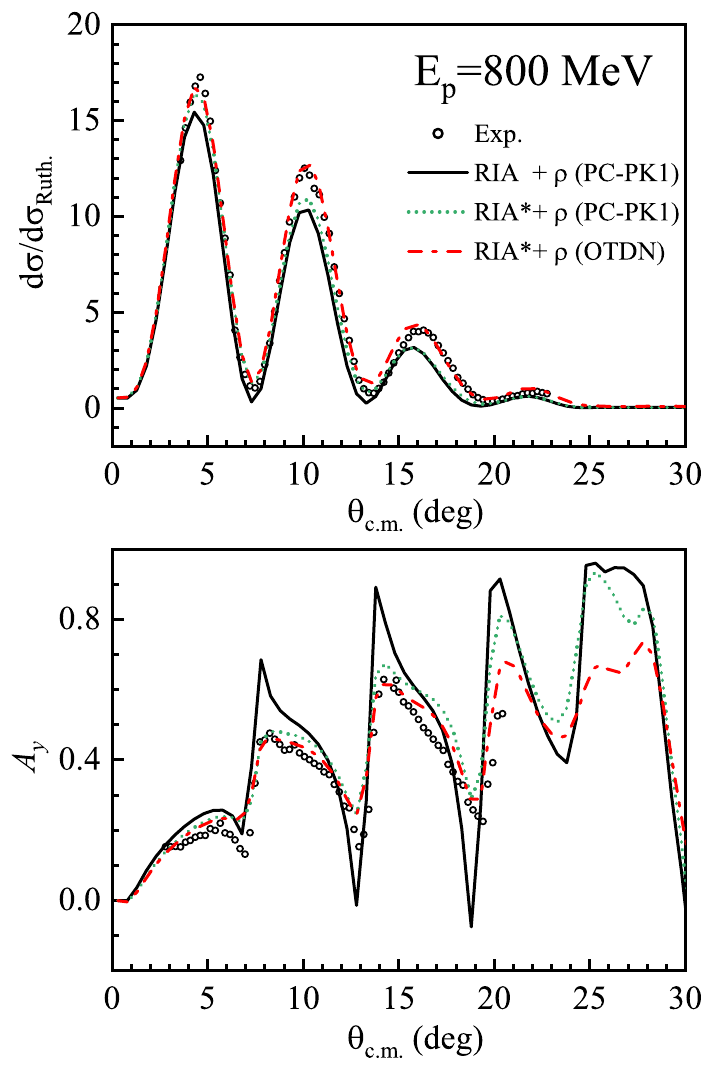}
  \caption{\label{fig:epsart3}
  The variation of the proton elastic scattering differential cross section $d\sigma/d\sigma_\text{Ruth}$ (upper panel) and analyzing power $A_{y}$ (lower panel) of $\mathrm{^{48}Ca}$ at 800 MeV as a function of the scattering angle $\theta_\text{c.m.}$. 
  The black solid line represents the theoretical results derived from the energy-dependent $NN$ interaction (Maxwell parameter set \cite{MAXWELL1998747}) with the PC-PK1 functional densities, the red dashed line marks the optimized $NN$ interaction with the PC-PK1 functional densities, and the green dotted line obtained based on the optimized $NN$ interaction with the OTDN-extracted densities.}
\end{figure}

As a further verification, the OTDN densities are input to RIA for recalculating the center-of-mass proton elastic scattering observables of $\mathrm{^{48}Ca}$ at an incident energy of 800 MeV--the differential cross section $d\sigma/d\sigma_\text{Ruth}$ and analyzing power $A_{y}$, which are illustrated in Fig.~\ref{fig:epsart3}.
Physically, these observables are determined by the angular distribution $\theta_\text{c.m.}$ of scattering amplitudes. 
Empirically, changes at small scattering angles are often associated with the state of the nuclear surface, while the information at larger scattering angles provides feedback on the nuclear internal properties.
The black solid line represents the theoretical results derived from RIA with the PC-PK1 functional densities, the red dashed line marks the RIA* with the PC-PK1 functional densities, and the green dotted line obtained based on the RIA* with the OTDN-extracted densities.

In comparisons, it can be observed that the optimized $NN$ interaction provides a more accurate description of the ``peaks" and ``valleys" in the differential cross section for $\theta_\text{c.m.}<15^\circ$. 
In terms of analyzing power, the results overall show better agreement with experimental values. 
This indicates that, at the current energy, the optimized interaction better captures the elastic scattering dynamics.
For broader applications in the future, ongoing efforts are necessary for further optimization of $NN~t$-matrix parameters.
Furthermore, introducing the optimized OTDN densities, the experimental values are excellently reproduced via RIA*-based calculations, particularly for the differential cross section. 
It must be emphasized that the density-induced observables change significantly at relatively large angles, while it remains relatively unchanged at smaller angles, corresponding to the optimization of density distribution primarily occurring inside nucleus (see Fig.~\ref{fig:epsart2}).

 \begin{figure}[tb]
  \includegraphics[width=8.5 cm]{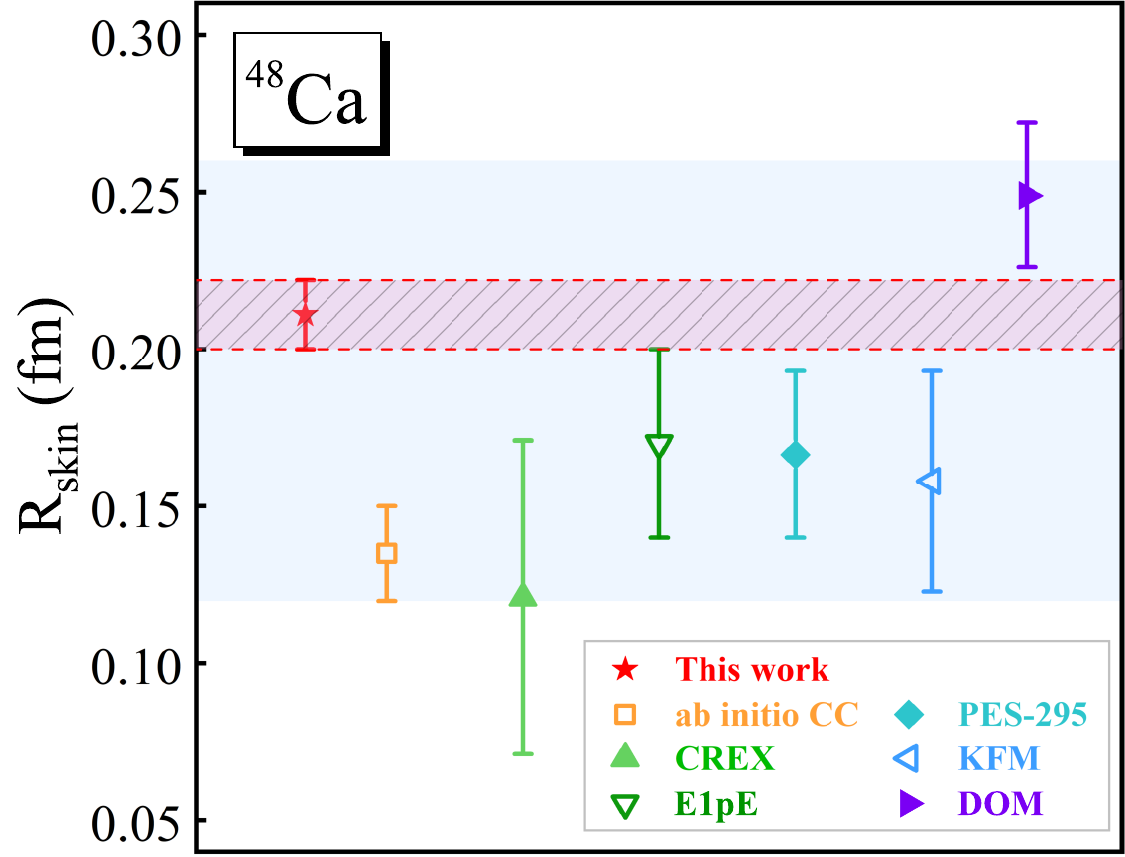}
  \caption{\label{fig:epsart4}
  Comparison of neutron skin thickness for $\mathrm{^{48}Ca}$ extracted from theories and experiments.
  The theoretical calculations and experimental measurements encompass 
  1.~The OTDN extracted results from this work. 
  2.~The first-principles $ab~~initio$ coupled-cluster (CC) methods \cite{30b9b0d97b934ea4bf5a7c9f73717cae}; 
  3.~The experimental measurement from CREX collaboration \cite{PhysRevLett.129.042501}; 
  4.~The high-resolution E1 polarizability experiment (E1pE) \cite{PhysRevLett.118.252501};
  5.~The phenomenological RIA-based derivation of proton elastic scattering at 295 MeV (PES-295) \cite{Zenihiro2018DirectDO};
  6.~The Kyushu (chiral) g-matrix folding model (KFM) for $AA$ and $pA$ scattering \cite{e72caed11f9c4f80bec9086cea8cc196};
  7.~The dispersion optical model (DOM) analysis \cite{PhysRevLett.119.222503};
  8.~Microscopic theoretical calculations from 48 reasonable energy density functionals (EDFs) \cite{PhysRevC.85.041302,PhysRevLett.119.222503} marked with the cyan shadow.}
\end{figure}

Considering the the excellent reproduce for the observables, the densities derived by OTDN appear to be reliable.
The final neutron skin thickness of $\mathrm{^{48}Ca}$ is
\begin{equation}
    R_\text{skin,final} = 0.211(11)\, \text{fm}
 \end{equation}
with neutron radius being $R_{n,\text{final}} = 3.604(6)\,\text{fm}$, and proton radius being  $R_{p,\text{final}}=3.393(5)\, \text{fm}$.
The corresponding comparisons with other theoretical calculations and experimental measurements are illustrated in Fig.~\ref{fig:epsart4}.
Our calculations, utilizing prior information provided by experiments and employing BMA to minimize theoretical approximations and eliminate model dependency, are reliable to a great extent.
It falls within the range of the 48 EDFs, exceeding values from other theories and measurements except for the DOM analysis.
A larger neutron skin thickness implies a greater symmetry energy slope, corresponding to a harder symmetry energy variation. 
This result has far-reaching implications for astrophysics and nuclear reactions.
Considering the linear relation between the symmetry energy slopes and neutron skin thicknesses at the saturation density point, the recent measurement for $^{208}$Pb with a thick neutron skin \cite{PhysRevLett.126.172502,PhysRevLett.126.172503} precisely demands that the neutron skin thickness for $^{48}$Ca is greater than 0.19 fm. 
The response relation has been proved by hundreds of energy density functional calculations \cite{e72caed11f9c4f80bec9086cea8cc196}.
We look forward to further experimental validation.

The relevance retrieved from hundreds of pseudodensities and observables indeed facilitates to refine nuclear density distribution. 
As superiority, due to the employed EDFs covering a larger range of symmetry energy slopes, the resulting density is no longer model-dependent on the EDF.
Furthermore, it can be observed that ensemble learning based on BMA has led to densities converging to a narrow range, indicating a smaller neutron skin error. 
The current approach holds significant potential for application to other spherical nuclei, such as $^{208}$Pb. 
However, it must be emphasized that there is still room for optimization, particularly in some approximations related to RIA.
For example, the RIA theory, when emulating the $NN$ interaction, does not account for multiple collisions, which may be disadvantageous for describing elastic scattering reactions involving heavy nuclei.

\section{SUMMARY}

This work novelly combines RIA and deep learning methods to infer the density distribution and neutron skin thickness of $\mathrm{^{48}Ca}$.
Initially, 600 pseudodensities are generated by randomly combining density calculations obtained through 16 EDFs covering a wide range of symmetry energy slopes.
Applying optimized RIA, the pseudodensity-corresponded observables are obtained, which are differential cross sections and analyzing powers.
These data are utilized to train a carefully designed deep neural network--OTDN, aiming to establish the inverse mapping from observables to density.
After inputting the experimental data into OTDN, the densities are calibrated.

In further validation, it is observed that the optimization of interactions and the generated density distribution collectively a positive impact on the observables, bringing them closer to experimental values.
The optimization of density primarily occurs internally, leading to a more noticeable improvement in the observables at large scattering angles.
Based on this, the neutron thickness is calculated to be 0.211(11) fm.
This relatively thicker result is deemed reasonable from the perspective of density functional analysis.
This approach not only significantly reduces the dependency on various  EDFs for nuclear density, but also minimizes the uncertainties from the approximations on density functional theory.

\begin{acknowledgments}

The authors thank Dr.~Sibo Wang for valuable discussions and communications.  
This work is supported partly by the National Natural Science Foundation of China (Grants No. 12375126, No. 12205030) and the Fundamental Research Funds for the Central Universities. 

\end{acknowledgments}

\nocite{1}

\bibliography{ref}

\end{document}


\preprint{APS/123-QED}

\title{Supplemental Material for“Extracting neutron skin from elastic proton-nucleus scattering with deep neural network”}

\author{G. H. Yang} \thanks{These authors contributed equally to this work.}
\affiliation{School of Physical Science and Technology, Southwest University, Chongqing 400715, China}

\author{Y. Kuang} \thanks{These authors contributed equally to this work.}
\affiliation{School of Physical Science and Technology, Southwest University, Chongqing 400715, China}

\author{Z. X. Yang}\email[zuxing.yang@riken.jp]{}
\affiliation{RIKEN Nishina Center, Wako 351-0198, Japan}
\affiliation{School of Physical Science and  Technology, Southwest University, Chongqing 400715, China}

\author{Z. P. Li}\email[zpliphy@swu.edu.cn]{}
\affiliation{School of Physical Science and Technology, Southwest University, Chongqing 400715, China}

\date{\today}

\begin{abstract}

This supplemental material includes two parameter tables, which are Tab.~\ref{tab:table1} for Relativistic Love-Franey NN interaction parameters at 800 MeV , and Tab.~\ref{tab:table2} for the feedforward hyperparameters of observable-to-density network (OTDN).

\end{abstract}

\maketitle


\section{Relativistic Love-Franey NN interaction parameters at 800 MeV}\label{A}

The following parameters are determined using the Levenberg-Marquardt method. 
The specific fitting procedure is consistent with Ref.~\cite{Li2007EnergydependentLC}, where the energy dependence is not taken into account.

\begin{table*}[th]
  \caption{\label{tab:table1}Real and imaginary Relativistic Love-Franey NN interaction parameters. 
  The masses $m$ and cutoff parameters $\Lambda$ are in MeV, whereas the coupling constants $g$ are dimensionless.}
  \begin{ruledtabular}
    \renewcommand\arraystretch{1.4}{
    \begin{tabular}{ccccccccccc}
      &\multicolumn{6}{c}{Real parameters}                                      \\\hline
        &  Meson                & Isospin & Coupling type    & m    &  $g^2$  &  $\Lambda$    \\ \hline
        & $ \sigma$               & 0       & Scalar (S)       & 650  & -8.3320  & 771.6 \\
        & $\omega $               & 0       & Vector (V)       & 782  & 7.0920   & 803.4 \\
        &  $ t_{0} $               & 0       & Tensor (T)       & 1400 & -0.6031 & 1486.0  \\
        &  $ a_{0} $                & 0       & Axial vector (A) & 1200 & -0.5023 & 3488.0  \\
        &   $ \eta $                    & 0       & Pseudoscalar (P) & 450  & -14.32  & 450.1 \\
        &  $ \delta $               & 1       & Scalar (S)       & 500  & -0.3646 & 4041.0  \\
        & $ \rho $                 & 1       & Vector (V)       & 770  & 0.3485  & 1149.0  \\
        &  $t_{1} $                 & 1       & Tensor (T)       & 450  & 0.1044  & 595.6 \\
        &  $ a_{1} $                  & 1       & Axial vector (A) & 800  & $-6.525\times 10^{-2}$  & 820.0   \\
        &  $ \pi  $                  & 1       & Pseudoscalar (P) & 138  & 12.31   & 557.5 \\
        \hline
        & \multicolumn{6}{c}{Imaginary parameters}                                    \\
        \hline
        &  Meson                & Isospin & Coupling type    & $\overline{m}$   & $\overline{g}^2$      & $\overline{\Lambda}  $    \\ \hline
        & $ \sigma$             & 0       & Scalar (S)       & 1300 & -4.5500   & 1553.0  \\
        & $\omega $             & 0       & Vector (V)       & 700  & 2.2100    & 1479.0  \\
        &  $ t_{0} $            & 0       & Tensor (T)       & 550  & -0.1078 & 709.3 \\
        &  $ a_{0} $             & 0       & Axial vector (A) & 750  & -0.4360  & 751.0   \\
        &  $\eta $                & 0       & Pseudoscalar (P) & 500  & 7.4180   & 632.7 \\
        &  $ \delta $            & 1       & Scalar (S)       & 500  & 0.1295  & 743.0   \\
        & $ \rho $               & 1       & Vector (V)       & 500  & $-6.6280\times 10^{-3}$  & 531.5 \\
        &  $t_{1} $              & 1       & Tensor (T)       & 450  & $-1.1760\times 10^{-2}$  & 1160.0  \\
        &  $ a_{1} $             & 1       & Axial vector (A) & 850  & -0.1116 & 860.0    \\ 
        &  $ \pi  $             & 1       & Pseudoscalar (P) & 450  & 2.2470   & 1246.0 
       \end{tabular}}
    \end{ruledtabular}
  \end{table*}

\bibliography{ref}

\newpage

\section{The hyperparameters table for observable-to-density network}\label{B}

The following parameters provide a detailed display of the number of hidden layers, neurons, and activation function for the observable-to-density network (OTDN) feedforward schematic diagram. 
\begin{table}[th]
\caption{\label{tab:table2}The feedforward hyperparameters of observable-to-density network (OTDN). 
The ``$L$" represents the number of hidden layers in each cell, the ``Type" marks the input, output, and hidden linear layers of the cell, the ``$D$" denotes the number of neurons, and the ``$g(x)$" signifies the activation function. 
All concepts used are consistent with those in PyTorch \cite{paszke2019pytorch}}

\renewcommand\arraystretch{1.2}{
\setlength{\tabcolsep}{7pt}
\begin{tabular}{cllc} \hline \hline
\multicolumn{4}{l}{Cell-1}     \\ \hline 
$\quad$ $L$ &$\quad$ Type & $D$ & $g(x)$     \\  \hline
\multicolumn{1}{c}{\quad---} &$\quad$ $\mathrm{Input_{1}}$  & 
75 & \multicolumn{1}{c}{---}    \\
$\quad$1 &$\quad$ Linear & 128 & ReLU    \\
$\quad$2 &$\quad$ Linear & 256 & ReLU    \\
\multicolumn{1}{c}{\quad---} &$\quad$ $\mathrm{Output_{1}}$ & 256 & \multicolumn{1}{c}{---}     \\\hline
\multicolumn{4}{l}{Cell-2}    \\  \hline
$\quad$ $L$ &$\quad$ Type & $D$ & $g(x)$     \\ \hline
\multicolumn{1}{c}{\quad---} &$\quad$ $\mathrm{Input_{2}}$ & 69 & \multicolumn{1}{c}{---}      \\
$\quad$1 &$\quad$ Linear & 128 & ReLU     \\
$\quad$2 &$\quad$ Linear & 256 & ReLU     \\
\multicolumn{1}{c}{\quad---} &$\quad$ $\mathrm{Output_{2}}$ & 256 & \multicolumn{1}{c}{---}     \\ \hline
\multicolumn{4}{l}{Cell-3}     \\ \hline
$\quad$ $L$ &$\quad$ Type & $D$ & $g(x)$    \\  \hline
\multicolumn{1}{c}{\quad---} &$\quad$ $\mathrm{Output_{1}}$ $\uplus$ $\mathrm{Output_{2}}$ & 512 & \multicolumn{1}{c}{---}     \\
$\quad$1 &$\quad$ Linear & 768 & ReLU    \\
$\quad$2 &$\quad$ Linear & 1024 & ReLU     \\
\multicolumn{1}{c}{\quad---} &$\quad$ $\mathrm{Output_{3}}$ & 1024 & \multicolumn{1}{c}{---}     \\ \hline
\multicolumn{4}{l}{Cell-4}     \\\hline
$\quad$ $L$ &$\quad$ Type & $D$ & $g(x)$     \\\hline
\multicolumn{1}{c}{\quad---}  &$\quad$ $\mathrm{Output_{3}}$  & 1024 & \multicolumn{1}{c}{---}     \\
$\quad$1 &$\quad$ Linear & 512 & ReLU     \\
$\quad$2 &$\quad$ Linear & 256 & ReLU     \\

$\quad$3 & $\quad$ Linear & 160 & Sigmoid     \\
\multicolumn{1}{c}{\quad---} &$\quad$ $\mathrm{Output_{4}}$ & 160 & \multicolumn{1}{c}{---}     \\  \hline
\multicolumn{4}{l}{Cell-5}     \\   \hline
$\quad$ $L$ &$\quad$ Type & $D$ & $g(x)$     \\   \hline
\multicolumn{1}{c}{\quad---} &$\quad$  $\mathrm{Output_{3}}$ & 1024 & \multicolumn{1}{c}{---}     \\
$\quad$1 &$\quad$ Linear & 512 & ReLU     \\
$\quad$2 &$\quad$ Linear & 256 & ReLU     \\
$\quad$3 &$\quad$ Linear & 160 & Sigmoid     \\
\multicolumn{1}{c}{\quad---} &$\quad$ $\mathrm{Output_{5}}$ & 160 & \multicolumn{1}{c}{---}     \\   \hline
\multicolumn{2}{l}{Other hyperparameters} & \multicolumn{2}{l}{Values and Properties}     \\   \hline
\multicolumn{2}{l}{Numerical Normalization Factor} & \multicolumn{2}{l}{10}    \\   
\multicolumn{2}{l}{Loss Function} & \multicolumn{2}{l}{NPD}     \\
\multicolumn{2}{l}{Optimizer} & \multicolumn{2}{l}{Adam}     \\
\multicolumn{2}{l}{Epoch 0-1000} & \multicolumn{2}{l}{lr = $ 1\times 10^{-2}$}   \\
\multicolumn{2}{l}{Epoch 1000-2000} & \multicolumn{2}{l}{lr = $1\times 10^{-3}$ }    \\ \hline \hline   
\end{tabular}} 
\end{table}